\documentclass[journal]{IEEEtran}
\usepackage{graphicx}

\ifCLASSINFOpdf
   \DeclareGraphicsExtensions{.pdf,.jpeg,.png}
\else
\fi

\newcommand{\beq}{\begin{equation}}
\def\eeq{\end{equation}}
\newcommand{\be}{\begin{enumerate}}
\newcommand{\ee}{\end{enumerate}}
\newcommand{\bi}{\begin{itemize}}
\newcommand{\ei}{\end{itemize}}
\newcommand{\bc}{\begin{center}}
\newcommand{\ec}{\end{center}}

\usepackage{tabularx}
\usepackage{multirow}
\usepackage{booktabs}
\usepackage{color}

\begin{document}
\title{Energy Spectrum Extraction and Optimal Imaging via Dual-Energy Material Decomposition}

\author{Wei Zhao,~Lu Wan,~Bo Zhang,~Qiude Zhang,~Zhangjing Xiong,~Tianye~Niu
\vspace{-1mm}
\thanks{Manuscript received May 3, 2015. This work was supported in part by the NSFC under Grant No. 61425001, 61210003, U1201256, by the National Key Scientific Instrument and Equipment Development Project of China under Grant No. 2013YQ030923, by the International S$\&$T Cooperation Program of China (ISTPC) under Grant No. 2014DFR10670, and by the Key Project of Chinese Ministry of Education under Grant No. 313023.       }
\thanks{W. Zhao is with the Wuhan Jiubang Technology Co., Ltd, Wuhan, Hubei 430074 China~(e-mail: zhaow85@163.com).}
\thanks{L. Wan, and Q. Zhang are with the Wuhan Riverine Technology Co., Ltd, and also with the Department of Biomedical Engineering, Huazhong University of Science and Technology, Wuhan, Hubei 430074 China.}
\thanks{B. Zhang is with the Wuhan digital PET Technology Co., Ltd, Wuhan, Hubei 430074 China.}
\thanks{T. Niu is with the Sir Run Run Shaw Hospital, Institute of Translational Medicine, Zhejiang University, Hangzhou, China~(e-mail: tyniu@zju.edu.cn).}
}

\maketitle

\begin{abstract}
Inferior soft-tissue contrast resolution is a major limitation of current CT scanners. The aim of the study is to improve the contrast resolution of CT scanners using dual-energy acquisition. Based on dual-energy material decomposition, the proposed method starts with extracting the outgoing energy spectrum by polychromatic forward projecting the material-selective images. The extracted spectrum is then reweighted to boost the soft-tissue contrast. A simulated water cylinder phantom with inserts that contain a series of six solutions of varying iodine concentration (range, 0-20 mg/mL) is used to evaluate the proposed method. Results show the root mean square error (RMSE) and mean energy difference between the extracted energy spectrum and the spectrum acquired using an energy-resolved photon counting detector(PCD), are 0.044 and 0.01 keV, respectively. Compared to the method using the standard energy-integrating detectors, dose normalized contrast-to-noise ratio (CNRD) for the proposed method are improved from 1 to 2.15 and from 1 to 1.88 for the 8 mg/mL and 16 mg/mL iodine concentration inserts, respectively. The results show CT image reconstructed using the proposed method is superior to the image reconstructed using the standard method that using an energy-integrating detector.
\end{abstract}


\section{Introduction}
%
%
%
%
\IEEEPARstart{C}{omputed} tomography (CT) is one of the most important imaging modalities for disease diagnosis, management and treatment. However, a major limitation of the current CT scanners is the low soft-tissue contrast resolution. Specifically, while bone, lung and tissue can be well differentiated from each other as their attenuation coefficients are significant different, organs and tissues that have similar attenuation coefficients are not well differentiated (such as renal mass), indicating relatively limited contrast. While keeping the imaging dose level constant, the limited contrast yields poor contrast-to-noise ratio (CNR), which reduces detectability of lesion in the internal organs. In order to increase CNR of the CT image, one may want to increase the radiation dose, however, the radiation dose to the patient must not exceed acceptable limits.

To overcome this limitation and to improve the clinical significance of CT scanners, novel imaging methods such as phase-contrast imaging (PCI) which generates radiographic contrast from the phase shift of x-rays passing through the object~\cite{davis1995,pfeiffer2006}, and energy-resolved CT imaging using photon-counting detectors (PCDs)~\cite{shikhaliev2008,schmidt2009,wang2011} have been extensively investigated recently. Even though PCI and PCD have shown superior contrast in CT images, neither PCI or PCD has been used in clinical diagnosis until now. Compared with X-ray imaging based on attenuation, X-ray imaging based on PCI is generally considered to be more sensitive to the mechanical vibration, with limited field-of-view, higher dose and longer time for data acquisition. For CT imaging using PCDs, although its performance has been significantly improved in the past decade, PCDs (such as CdTe detector) with energy discrimination based on pulse analysis are still suffered from limited count rate which can not fully meet the requirement of clinical applications. In addition, PCDs usually have higher cost and their uniformity, long-term reliability and stability still needs to be improved for clinical CT applications. Due to these reasons, neither PCI or PCD has been used in clinical diagnosis.


For an energy integrating detector (EID), the detector output of a pixel is proportional to the energy deposition in this pixel. Thus high energy photons that provide lower contrast contribute more to the measurement, while low energy photons that provide higher contrast contribute less to the measurement, yielding inferior weighted projection data. A major advantage of PCD against EID is the capability of energy discrimination, based on which optimal energy weights can be assigned to each photon to yield the best CNR image. Since dual energy material decomposition can yield material-specific images, it is possible to extract the energy spectrum information using polychromatic reprojection if the incident spectrum is well modeled.

In this study, we demonstrate how the dual-energy material decomposition technique can be used to extract x-ray energy spectrum from projection data acquired using EIDs. Furthermore, the extracted outgoing polychromatic energy spectrum is optimized to boost soft-tissue contrast, as that was done in the energy-resolved PCD-based CT imaging.

\section{Methods and materials}
\begin{figure}[t]
    \centering
    \includegraphics[width=3.5in]{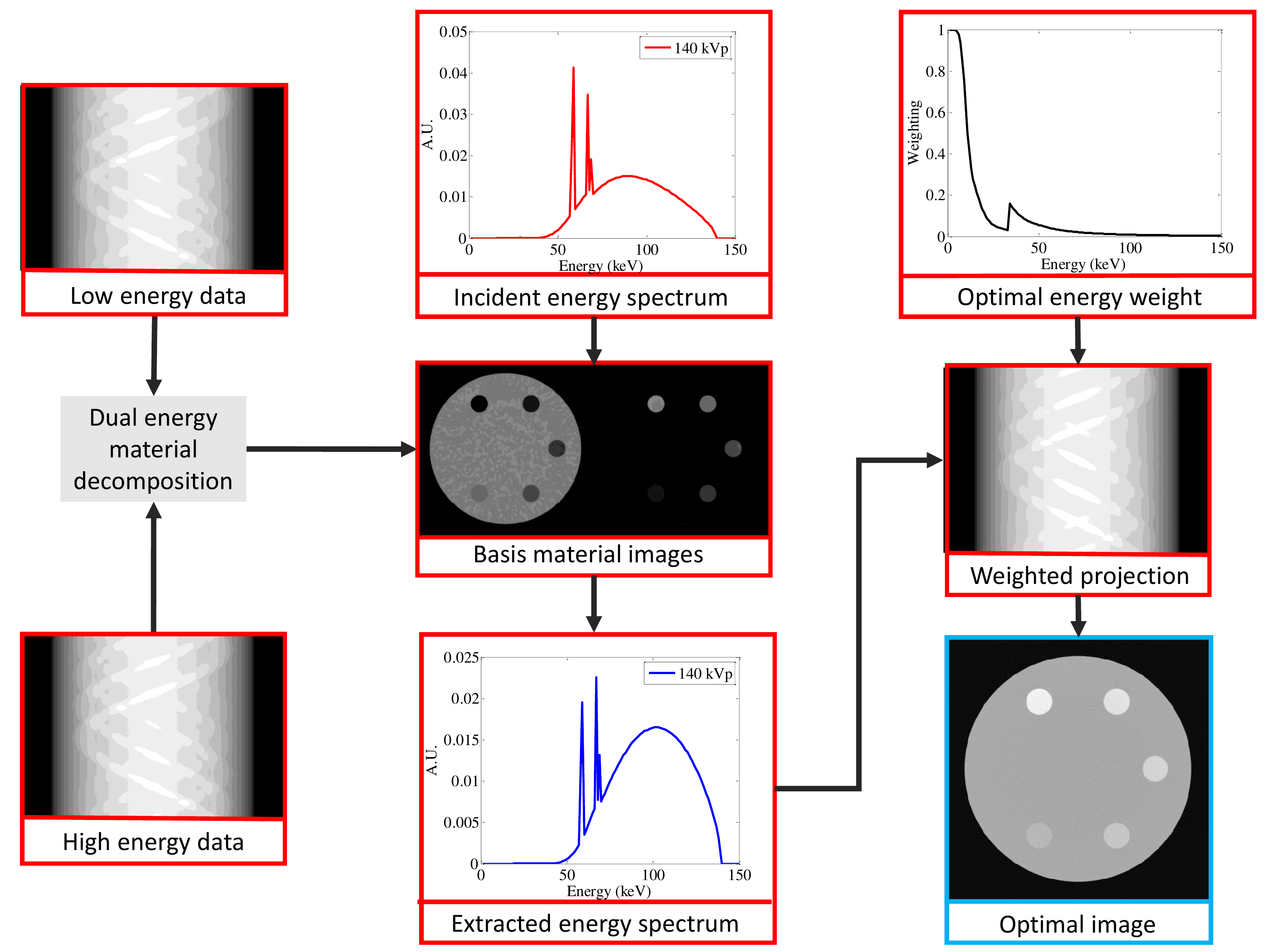}
    \vspace{-1em}
    \caption{A flowchart of the proposed polychromatic energy spectrum extraction and optimal imaging method. }
    \vspace{-1.5em}
    \label{fig:f1}
\end{figure}

Figure~\ref{fig:f1} shows the flowchart of the proposed polychromatic energy spectrum extraction and optimal imaging method. The method starts with dual-energy material decomposition to obtain the material-specific images, based on which energy spectrum of each detector pixel is extracted using polychromatic reprojection. The energy information of the spectrum is then reweighted using precalculated energy weighting, yielding a set of optimized projection data. CNR improved image can be obtained by reconstructing the optimized projection. The rationale of the method is that by introducing accurate material decomposition, the energy component and the location component of the attenuation coefficient is decoupled and its energy-dependent property has been transferred into the basis materials. Hence if the incident spectrum is precisely modeled, the outgoing spectrum can be calculated with the material-selective images and the known basis materials. In the following subsections, we introduce the three major components of the proposed method, i.e., dual-energy material decomposition, polychromatic reprojection, and energy extraction and optimization.

\subsection{Dual-energy material decomposition}
Dual-energy material decomposition based on the assumption that the linear attenuation coefficient $\mu(\vec{r},E)$ can be modeled as a weighted summation of two other materials,
\beq\label{equ:decomposition}
\mu(\vec{r},E)=f_{1}(\vec{r})\psi_{1}(E)+f_{2}(\vec{r})\psi_{2}(E)
\eeq
Here $\psi_{1,2}$ are the known independent energy dependencies which can be mass attenuation coefficients of basis materials and $f_{1,2}(\vec{r})$ are the material-selective images. Material decomposition can be performed in either projection domain or image domain. For both of the cases, noise is amplified if no measurement is taken. To address the amplified noise issue, an iterative image domain material decomposition method that balances the data fidelity of the value of the material image and a quadratic penalty using an optimization framework, is employed in this work~\cite{niu2014}. The unconstrained optimization problem is solved by the nonlinear conjugate gradient method.

\subsection{Polychromatic reprojection}
Based on the material decomposition expression, polychromatic projection of an object is represented as
\beq\label{equ:polyreprojBimg}
\hat{I_i}=\int_{0}^{E_{max}}\mathrm{d}E\,\Omega(E) \, \eta(E)\,\mathrm{exp}\left[-A_{1i}\psi_{1}(E)-A_{2i}\psi_{2}(E)\right],
\eeq
with $A_{1i}=\int_{L_i}\mathrm{d}\vec{r}\,f_{1}(\vec{r})$ and $A_{2i}=\int_{L_i}\mathrm{d}\vec{r}\,f_{2}(\vec{r})$ are the line integrals of the material-selective images. Here $L_i$, $\Omega(E)$ and $E_{max}$ are the propagation path length of ray $i$, the corresponding polychromatic x-ray spectrum of the ray and the maximum photon energy of the spectrum, respectively. $\eta(E)$ is the energy dependent response of the detector.

\subsection{Energy extraction and optimization}

Eq~(\ref{equ:polyreprojBimg}) can be further reduced as
\beq\label{equ:spekEx}
\hat{I_i}=\int_{0}^{E_{max}}\mathrm{d}E\,\phi_i(E)\, \eta(E),
\eeq
with $\phi_i(E)=\Omega(E) \,\mathrm{exp}\left[-A_{1i}\psi_{1}(E)-A_{2i}\psi_{2}(E)\right]$. Since the incident spectrum $\Omega(E)$ can be obtained from either direct measurement or indirect transmission measurement~\cite{zhao2015}, we can calculate the outgoing spectrum of ray $i$ based on the material-specific images.

For an EID, $\eta(E)$ is proportional to the detected photon energy, which is the case for most of realistic CT applications and yields inferior CT images. To improve CNR of the CT images, it is possible to reweight the contribution of the detected photon by using the knowledge of the outgoing spectrum. The optimal energy weights $w(E)$ is calculated as~\cite{giersch2004},
\beq\label{equ:oWeight}
w(E)=\frac{1-\delta}{1+\delta},
\eeq
with $\delta=\mathrm{exp}\left[(\mu_b(E)-\mu_t(E))\,s\right]$, where $\mu_b$, $\mu_t$, and $s$ are the attenuation coefficients of the background material, the target material and the expected size of the target.
Thus the final optimized projection data is,
\beq\label{equ:oWeight}
\hat{I_i}^w=I_i^m\,\frac{\int_{0}^{E_{max}}\mathrm{d}E\,\phi_i(E)\, w(E)}{\int_{0}^{E_{max}}\mathrm{d}E\,\phi_i(E)\, \eta(E)}.
\eeq
Here $I_i^m$ is the measured projection data using the EID.

\subsection{Simulations}

\begin{figure}[t]
    \centering
    \vspace{-1em}
    \includegraphics[width=3.6in]{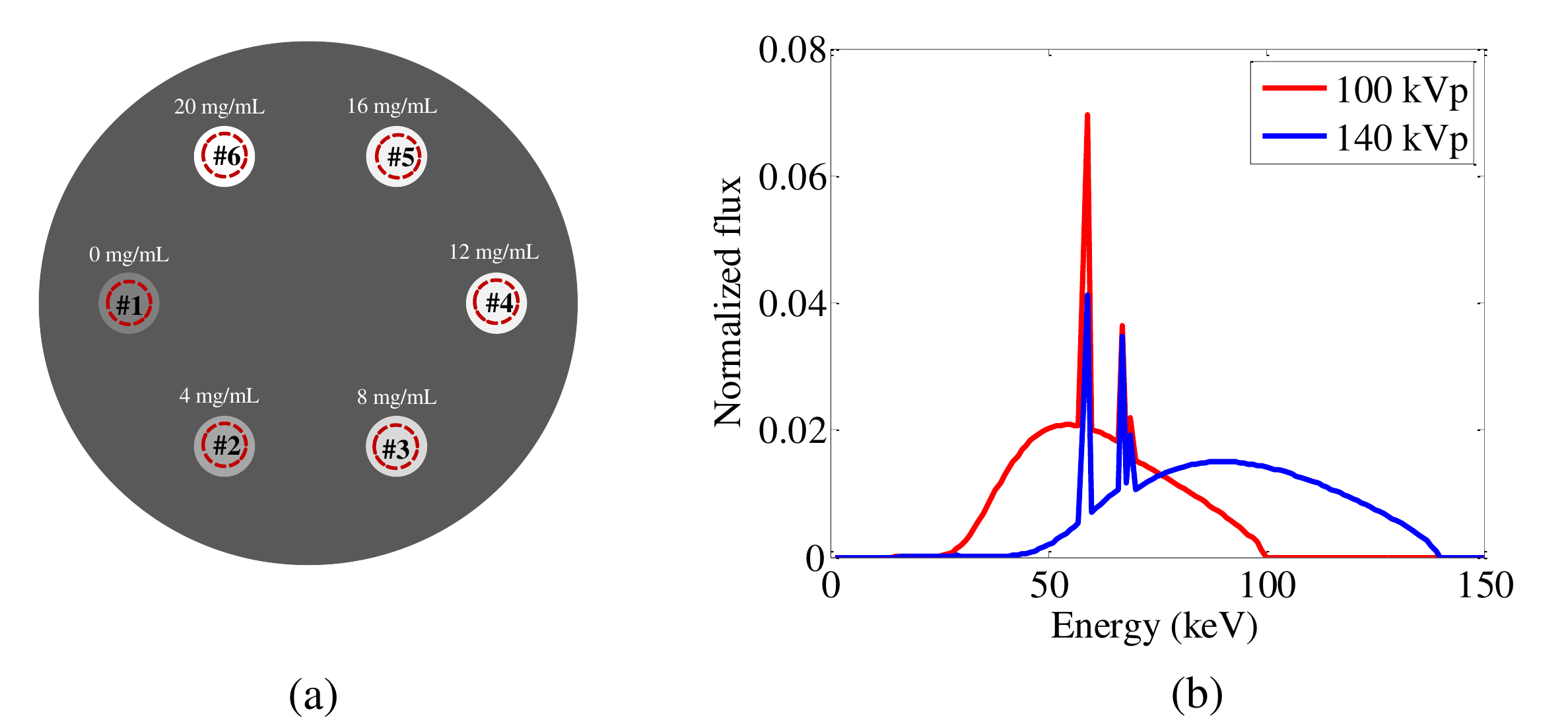}
    \caption{Iodine concentration phantom (a) and x-ray spectra (b) for numerical simulation. }
    \vspace{-0.8em}
    \label{fig:f2}
\end{figure}

\begin{figure}
    \centering
    \includegraphics[width=2.8in]{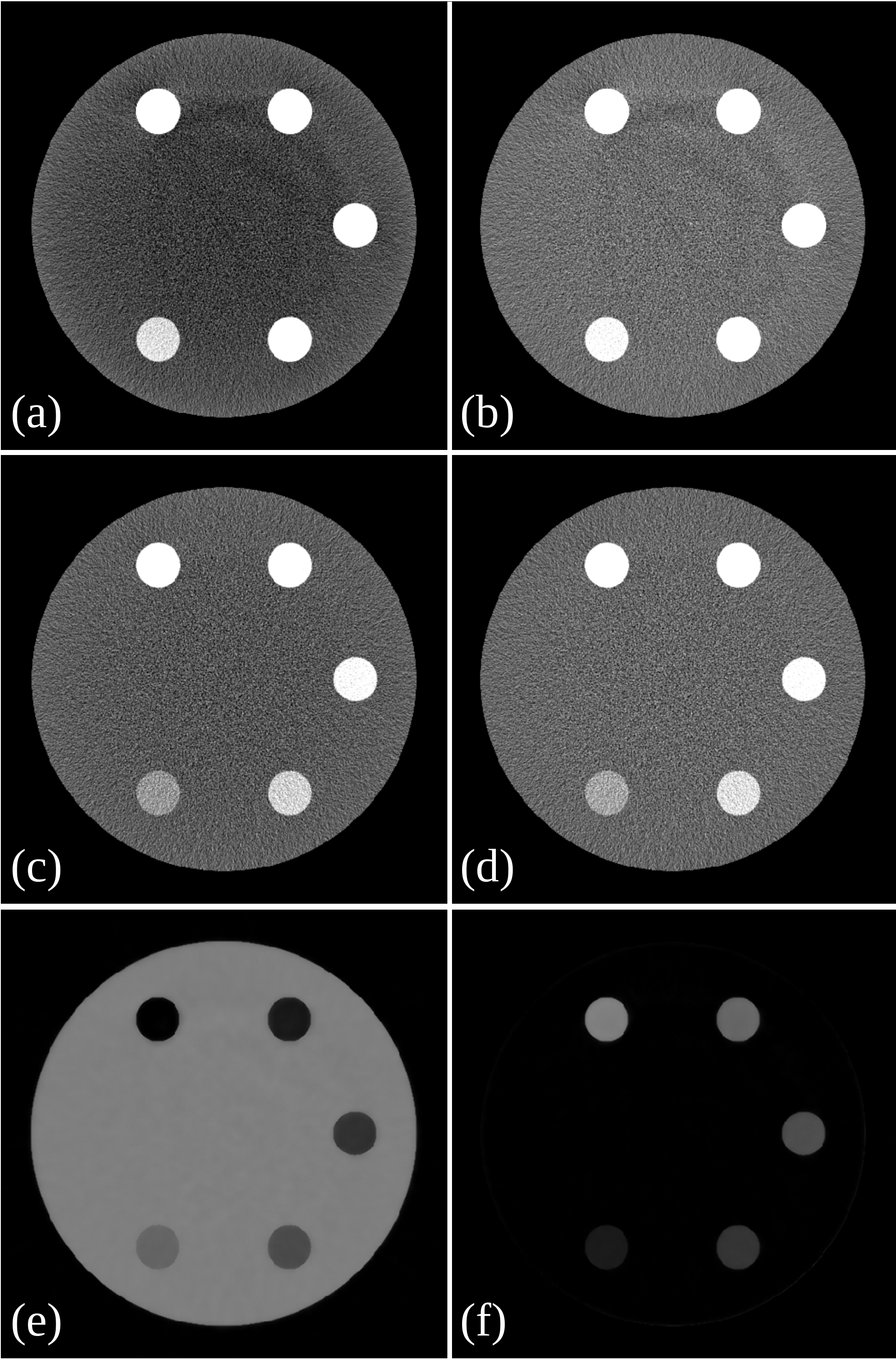}
    \caption{CT images and material-specific image for the numerical iodine concentrate phantom. 100 kV CT images before (a) and after (b) first order beam hardening correction (Display window: [0.019, 0.023] mm$^{-1}$). 140 kV CT images before (c) and after (d) first order beam hardening correction (Display window: [0.016, 0.020] mm$^{-1}$). Water image (e) and iodine image (f) generated using iterative image-domain dual-energy material decomposition (C/W = 0\%/200\%). }
    \label{fig:f3-1}
\end{figure}

A water cylinder phantom with inserts that contain a series of six solutions of varying iodine concentrate (range, 0-20 mg/mL in step of 4 mg/mL) is used to evaluate the proposed method. The energy spectra for dual energy CT imaging were 100 kVp and 140 kVp, and they were generated using the SpekCalc software~\cite{poludniowski2009} with 12 mm Al and 0.4 mm Sn + 12 mm Al filtration, respectively. During the simulation, Poisson noise was introduced and the first order beam hardening correction was performed. The x-ray spectra and the phantom were shown in figure~\ref{fig:f2}. Iterative image-domain dual-energy material decomposition method was employed to generate the material-specific images, which were applied to the material image-based polychromatic forward projection to extract the energy spectrum.

In order to quantitatively characterize the accuracy of the estimated spectrum, we use the normalized root mean square error NRMSE and the mean energy difference between the true spectrum and the estimated spectrum $\Delta E$ which were defined as follows:
\begin{equation}\label{equ:error}
NRMSE=\sqrt{\frac{\sum_{e=1}^{N}(\hat{\Omega}(e)-\Omega(e))^{2}}{\sum_{e=1}^{N}\Omega(e)^{2}}}    
\end{equation}
\begin{equation}\label{equ:meanEnergy}
\Delta E= \sum_{e=1}^{N}E(e)\,(\Omega(e)-\hat{\Omega}(e))
\end{equation}
Here $\hat{\Omega}(e)$ is the $e$th energy bin of the normalized estimated spectrum and $\Omega(e)$ is $e$th energy bin of the normalized raw spectrum. $N$ is the total number of the energy bins of the spectrum, and $E(e)$ is the energy of the $e$th energy bin.



\section{Results}

\begin{figure}
    \centering
    \includegraphics[width=3.5in]{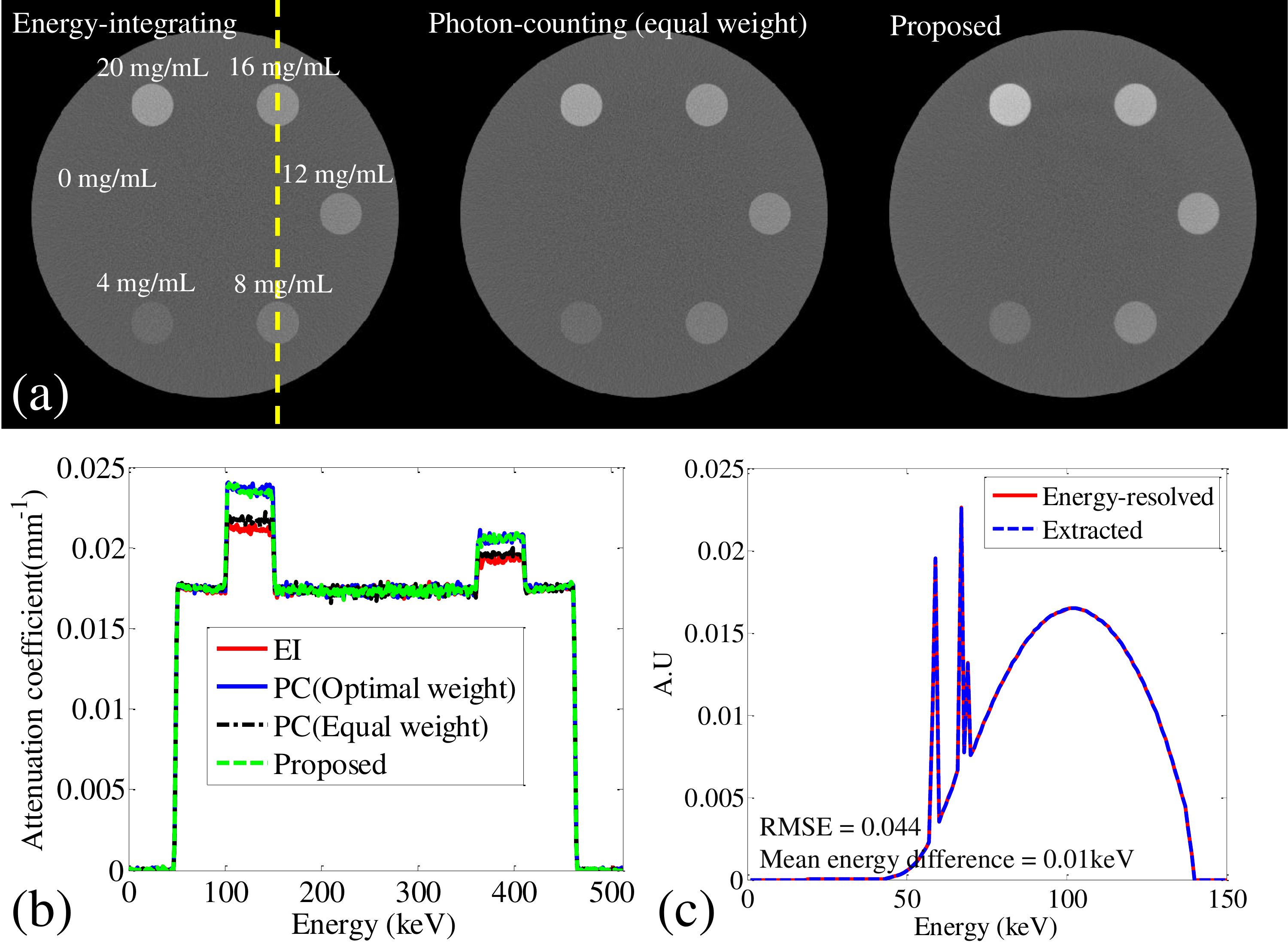}
    \caption{Results for the iodine concentrate phantom. (a) CT images reconstructed using projection data obtained from energy-integrating detector, photon counting detector, and proposed method. (b) Line profiles (labelled as the yellow line) of the CT images. (c) X-ray spectra acquired using energy-resolved detector and extracted using the proposed method. }
    \label{fig:f3}
\end{figure}

\begin{table}
\caption{Dose normalized contrast-to-noise ratio (CNRD) for the different methods. The CNRDs are normalized to the corresponding EID values.}
\label{tab:CNRD}
\begin{center}
\begin{tabular}{ccccc}
\toprule
\rule[-1ex]{0pt}{3.5ex}  Inserts  & EID & PCD& PCD& Proposed\\
(mg/mL) &  & (equal weight)& (optimized weight) & \\
\hline
\rule[-1ex]{0pt}{3.5ex}  8  & 1.00 & 1.42 & 1.74 & 2.15  \\
\rule[-1ex]{0pt}{3.5ex}  16  & 1.00 & 1.18 & 1.52 & 1.88  \\
\bottomrule
\end{tabular}
\end{center}
\vspace{-1mm}
\end{table}

Figure~\ref{fig:f3-1} shows CT images and material-specific images for the numerical iodine concentrate phantom. As can be seen, after first order beam hardening correction (water correction), cupping artifacts in the 100 kV CT image are almost completely removed, however, there are residual steaks (high order beam hardening artifacts) between the dense inserts, as illustrated in figure~\ref{fig:f3-1} (a) and (b). Beam hardening correction for the 140 kV CT image is less noticeable (figure~\ref{fig:f3-1} (c) and (d)) since the 140 spectrum is much harder. Figure~\ref{fig:f3-1} (e) and (f) show water and iodine images, respectively. 

Figure~\ref{fig:f3} shows the results for the iodine concentrate phantom using numerical simulation. Figure~\ref{fig:f3}(a) shows the CT images reconstructed using projection data obtained from EID, PCD with equal energy weighting and the proposed method. As can be seen, the iodine concentration inserts show improved contrast.
Figure~\ref{fig:f3}(b) shows image generated using the proposed method has similar contrast as that using PCD with optimal energy weighting. Quantified with root mean square error (RMSE) and mean energy difference $\Delta E$, figure~\ref{fig:f3}(c) shows the spectra extracted using the proposed method and the energy-resolved detector have almost the same profiles. Table~\ref{tab:CNRD} shows the dose normalized CNR of the 8 mg/mL and 16 mg/mL inserts for the different methods. 
Compared with the method using the standard EID, CNRD for the proposed method are improved from 1 to 2.15 and from 1 to 1.88 for the 8 mg/mL and 16 mg/mL inserts, respectively.

\section{Summary and discussions}
\label{sec:discussion}
Based on dual-energy material decomposition, we propose a method to extract energy spectrum using EIDs. By polychromatic forward projecting the material-selective images, the outgoing spectrum for each detector pixel can be precisely extracted. The extracted energy spectrum is then reweighted in projection domain to yield CNR improved CT images. Results show spectrum extracted using the proposed method matches spectrum acquired using the energy-resolved detectors quite well, and CT image reconstructed using the proposed method is superior to the image reconstructed using the standard method and is comparable to the image obtained using photon-counting detector with optimal weights. The method requires the incident spectrum to be modeled precisely. This limitation can be overcome using our previous proposed indirect spectrum estimation method or other spectrum estimation or measurement methods. 


\bibliographystyle{IEEEtran}
%

\end{document}